\documentclass[aps,prl,twocolumn,showpacs,groupedaddress]{revtex4-1}

\usepackage{graphics}

\usepackage{epsfig}
\usepackage{amsmath}
\usepackage{amssymb}
\usepackage{amsfonts}
\usepackage{ dsfont }

\usepackage[debug]{mciteplus}

\usepackage{color}
\usepackage{ulem}

\newcommand{\be}{\begin{equation}}
\newcommand{\ee}{\end{equation}}

\begin{document}


\hsize\textwidth\columnwidth\hsize\csname@twocolumnfalse\endcsname

\title{Spin relaxation anisotropy in a GaAs quantum dot}
\author{P. Scarlino$^1$, E. Kawakami$^1$, P. Stano$^{2,3}$, M. Shafiei$^1$, C. Reichl$^3$, W. Wegscheider$^3$ and L. M. K. Vandersypen$^1$}
\affiliation{$^1$Kavli Institute of Nanoscience, TU Delft, Lorentzweg 1, 2628 CJ Delft, The Netherlands\\
$^2$RIKEN Center for Emergent Matter Science, Wako-shi, Saitama 351-0198, Japan\\
$^3$Institute of Physics, Slovak Academy of Sciences, Dubravska cesta 9, 84511 Bratislava, Slovakia\\ 
$^4$Solid State Physics Laboratory, ETH Zurich, Schafmattstrasse 16, 8093 Zurich
}
\date{\today}
\vskip1.5truecm
\begin{abstract}
We report that the electron spin relaxation time, $T_1$, in a GaAs quantum dot with a spin-1/2 ground state has a 180 degree periodicity in the orientation of the in-plane magnetic field. This periodicity has been predicted for circular dots as due to the interplay of Rashba and Dresselhaus spin orbit contributions. Different from this prediction, we find that the extrema in the $T_1$ do not occur when the magnetic field is along the $[110]$ and $[1\overline{1}0]$ crystallographic directions. This deviation is attributed to an elliptical dot confining potential. The $T_1$ varies by more than an order of magnitude when rotating a 3 Tesla field, reaching  about 80 ms for the $\it{magic}$ angle. We infer from the data that in our device the sign of the Rashba and Dresselhaus constants are opposite.
\end{abstract}
\maketitle
The high control reached in the manipulation of a single electron spin in a semiconductor environment \cite{Hanson2007} is encouraging for future application of this natural two-level system for quantum computation technology. In GaAs, InAs and other III-V quantum dots it has been shown that this manipulation can be realized using exclusively electrical fields \cite{Nowack2007,Nadj-Perge2010}. Coupling of the electric field to the spins is mediated by the spin-orbit (SO) interaction naturally provided by the semiconductor host environment. The semiconductor environment also implies that the electron is intimately in contact with phonons, charge fluctuations and nuclear spins and those interactions are responsible for the relaxation and dephasing process of the electron spin.

\begin{figure}
\includegraphics[width=8.5cm] {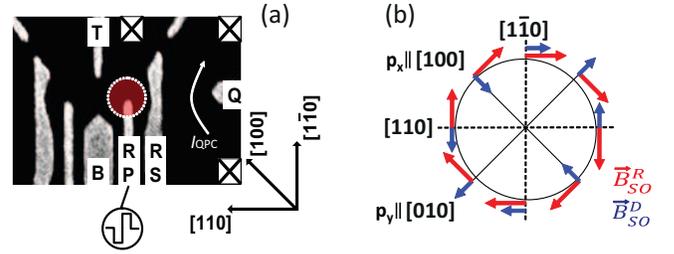}
\caption{\label{fig:fig1} (a) Scanning electron micrograph of a device similar to the one measured. The black arrows indicate the crystallographic axes. The dotted red circle represents schematically the single QD position. (b) The spin-orbit field ${\bf B}_{SO}$ acting on a conduction electron is shown by red and blue arrows, arising from the Rashba and Dresselhaus contribution respectively (chosen to be different in modulus and $\alpha<0$, $\beta>0$).}
\end{figure}

During the last ten years, a significant experimental \cite{Fujisawa2002,Elzerman2004,Kroutvar2004,Johnson2005,Amasha2006,Meunier2007,Amasha2008} and theoretical \cite{Khaetskii2000,Khaetskii2001,Golovach2004,Stano2006,Stano2006a} effort has been devoted to understanding the effect of electron spin relaxation in lateral quantum dots (QDs). At magnetic fields of the order of Tesla, spin relaxation in GaAs dots was found to be dominated by the SO interaction in combination with piezo-electric phonons.
Two contributions to the SO interaction usually dominate. The local electric field due to a crystal with bulk inversion asymmetry generates a Dresselhaus (D) SO contribution \cite{Dresselhaus1955} which, for electrons confined in the plane ($xy$, with $x$ and $y$ along the [100] and [010] crystallographic directions, respectively) of the quantum well, can be written as  $H_D=\beta(-\sigma_x P_x+\sigma_y P_y)/\hbar$, with $\hbar$ the Planck constant, $\beta$ the Dresselhaus SO coupling strength, ${\bf P}$ the electron kinematic momentum and $\boldsymbol{\sigma}$ the vector of Pauli matrices. In addition, the electric field associated with the asymmetric confining potential along the heterostructure growth direction ($z$ along [001]) gives rise to the Rashba (R) SO contribution \cite{Bychkov1984}, $H_R=\alpha(\sigma_y P_x-\sigma_x P_y)/\hbar$, with $\alpha$ the Rashba SO coupling strength. The effect of the SO interaction can be viewed as an effective magnetic field ${\bf B}_{SO}$ acting on the conduction electron spin, with an amplitude and direction that depend on the electron momentum [see Fig.~\ref{fig:fig1}(b)]. The interplay of R and D coupling gives rise to an anisotropy in the direction and magnitude of ${\bf B}_{SO}$ in the plane of the quantum well. As a result, spin relaxation in a quantum dot is anisotropic in the direction of the in-plane magnetic field \cite{Golovach2004,Stano2006}.

The anisotropy of electron spin relaxation originating from SO interaction has not been studied experimentally so far, even though it is highly relevant. Indeed, depending on the circumstances, it may be desirable to get long relaxation times or to make the relaxation process as fast as possible, for example in order to rapidly initialize the spin \cite{Srinivasa2013}. The SO anisotropy similarly affects the strength of the effective driving field for single-qubit rotations based on electric dipole spin resonance \cite{Golovach2006}. With a proper understanding, one can design future devices that optimally reconcile various requirements.

Here we present a measurement of $T_1$ as a function of the orientation of an in-plane magnetic field. We find a striking anisotropy with a 180 degree periodicity, confirming the theoretical predictions experimentally for the first time. Comparison with the predictions indicates that also the dot shape anisotropy plays an important role. We discuss in detail what information is needed to determine the ratio of the Rashba and Dresselhaus coupling strengths in this case. We also provide guidance for sample design and magnetic field orientation in future experiments. 

The experiment has been realized in a single depletion QD, see Fig.~\ref{fig:fig1}(a) created by applying a negative potential to surface gates on top of a GaAs/Al$_{0.33}$Ga$_{0.67}$As heterostructure, grown along the $[001]$ direction. The GaAs/AlGaAs interface is 85 nm deep, with Si-delta-doping of about 1.3$\times$10$^{-12}$cm$^{-2}$ atoms 40 nm away from the 2DEG, and a carrier density of 3.6$\times10^{11}$ cm$^2$/Vs. The base temperature of the dilution refrigerator was 25 mK and we estimated the electron temperature to be 130 mK from transport measurements at zero magnetic field. From pulse spectroscopy measurements \cite{Elzerman2004a} we infer that the dot contains most likely three electrons (see Supplemental Material \cite{suppl}, Sec.~I). Two electrons form a closed shell, with the third electron effectively acting as a spin-1/2 system. The orientation of the quantum dot gate pattern with respect to the main crystallographic directions is shown in Fig.~\ref{fig:fig1}(a). We applied a magnetic field in the 2DEG plane (at an angle $\phi$ with respect to the $[100]$ direction) of modulus 3 T, to ensure that the spin Zeeman energy ($\Delta_z\approx$ 60 $\mu$eV) is higher than the electron temperature ($k_BT_{el}\approx$15 $\mu$eV), as required for energy selective spin read-out (see below) \cite{Elzerman2004}. Real-time detection of the dot occupation is realized by monitoring the current through a quantum point contact (on the right side of the structure), amplified by a room temperature I-V converter, and low-pass filtered with a bandwidth of 30 kHz.

\begin{figure}
\includegraphics[width=8.5cm] {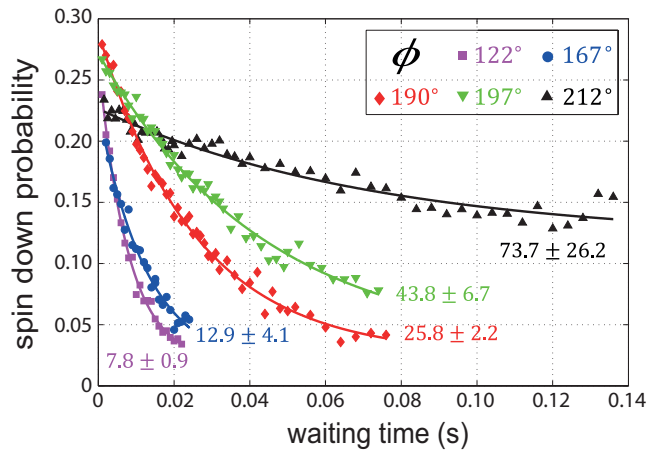}
\caption{\label{fig:fig2} Measured spin-down probability (averaged over ~5000 cycles) as a function of the waiting time between injection and read-out (see Supplemental Material \cite{suppl}, Sec.~I) for different angles $\phi$ of the 3 T in-plane magnetic field. The solid lines are fits to the data of the form $P_\downarrow = a \exp(-t/T_1) + b$. The fitted $T_1$'s are indicated for each curve (in ms). Small variations in $P_\downarrow(t=0)$ can arise from variations in the read-out configuration in the course of the measurements. Measuring longer $T_1$'s requires longer waiting times, with increased pulse distortion from the bias-tee (see Supplemental Material \cite{suppl}, Sec.~I), and therefore larger error.}
\end{figure}

\begin{figure}
\includegraphics[width=8.5cm] {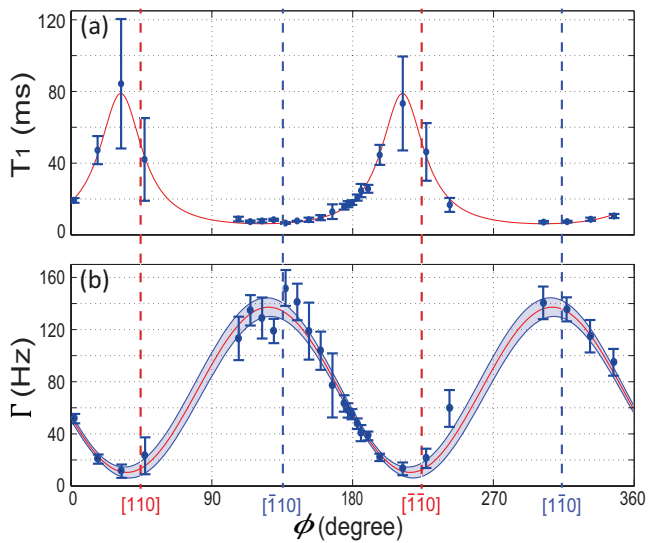}
\caption{\label{fig:fig3} Angle dependence of the spin relaxation time (a) and rate (b), which are separately extracted from exponential fits with either the relaxation rate or time in the exponent. The magnetic field is nearly in-plane, with $\left|\xi\right| < 5^\circ$, while the in-plane angle $\phi$ has a systematic error of $\pm 3^\circ$. The red line is a fit to Eq.~\eqref{eq:elongated} with free parameters $(\phi_{min},\xi^\ast,b)$. The shaded region between the two blue curves indicates the $95\%$ confidence interval. The dashed vertical lines show the positions of the extrema of $\Gamma$ predicted for a circular dot (at $\phi=45^\circ$ and $135^\circ$, see Eq.~\eqref{eq:circular 1}).}
\end{figure}

We measure the electron spin relaxation time by applying a three or four-stage pulse to gate RP \cite{Elzerman2004} (see also Supplemental Material \cite{suppl}, Sec.~I). 
The main observation is a striking variation in the relaxation time upon rotation of the in-plane magnetic field (Fig.~\ref{fig:fig2}).
Fig.~\ref{fig:fig3} shows the measured relaxation time as a function of the magnetic field orientation over the whole 360 degree range. The data shows clearly the predicted 180 degree periodicity and a remarkable variation in $T_1$ from 7 to 85 ms [Fig.~\ref{fig:fig3}(a)]. The maxima in $T_1$ are sharply peaked. When plotting the same data inverted, as $\Gamma=1/T_1$ [Fig.~\ref{fig:fig3}(b)], we see a sinusoidal variation of the relaxation rate. 

To understand this sinusoidal modulation, it is useful to express the spin relaxation rate in terms of a cross-product of the external field ${\bf B}=B(\cos\xi \cos\phi, \cos\xi \sin\phi, \sin\xi)$, and the in-plane vector 
\cite{Levitov2003}
\begin{equation}
{\bf n} =x \left( l_{d}^{-1}, -l_{r}^{-1},0\right) + y \left( l_{r}^{-1}, -l_{d}^{-1},0 \right),
\label{eq:nso}
\end{equation}
which refers to crystallographic directions $\hat{x}=[100]$, and $\hat{y}=[010]$ through the operator of electron coordinates ${\bf r}=(x,y)$. The SO lengths, $l_{r,d}=\hbar^2/(2m^\star\alpha,\beta)$, with $m^\star$ the effective electron mass, are defined as the distance travelled by an electron over which its spin is rotated by $\pi$ due to ${\bf B}_{SO}$ (typically 1-10 $\mu$m in GaAs). 
For a circular dot, the relaxation rate is (see Supplemental Material \cite{suppl}, Sec.~II)
\begin{equation}
\Gamma \propto |{\bf B} \times ( l_{d}^{-1}, -l_{r}^{-1},0 ) |^2 + |{\bf B} \times ( l_{r}^{-1}, -l_{d}^{-1},0 ) |^2,
\label{eq:rate 1}
\end{equation}
since the dipole operators $x$ and $y$ contribute equally \cite{Stano2006}. Parametrizing the SO lengths by $l_r^{-1}=l_{so}^{-1} \cos \vartheta$, and $l_d^{-1}=l_{so}^{-1}\sin\vartheta$, a straightforward evaluation of Eq.~\eqref{eq:rate 1} gives the known result \cite{Golovach2004,Stano2006}
\begin{equation}
\Gamma \propto l_{so}^{-2} \left[\sin^2 \xi + \cos^2 \xi (1+ \sin 2\phi \sin 2\vartheta) \right],
\label{eq:circular 1}
\end{equation}
which, for an in-plane magnetic field ($\xi=0$) reduces to
\begin{equation}
\Gamma \propto l_{so}^{-2} \left(1+ \sin 2\phi \sin 2\vartheta \right).
\label{eq:circular 2}
\end{equation}
For positive relative sign of the SO couplings, the rate reaches a maximum (minimum) with the external field along $[110]$ ($[1\overline{1}0]$). If the relative sign is inverted, the position of the minimum and maximum swap. If R and D have equal strength, the minimal rate is zero, while the sinusoidal modulation is reduced the more R and D differ in strength. Therefore, the relative strength of R and D, including the relative sign, can be extracted from the dependence of $\Gamma$ on the magnetic field orientation.

Looking at the data in Fig.~\ref{fig:fig3}, the extrema of the rate are shifted by $\approx 10^\circ$ from the prediction of Eqs.~\eqref{eq:circular 1} and \eqref{eq:circular 2}. Similar offsets were observed in the dependence of SO induced avoided level crossings on the magnetic field orientation in InAs dots \cite{Takahashi2010,Kanai2011}, and were explained by invoking anisotropic dot shapes \cite{Nowak2011}. The dot anisotropy influences also the spin relaxation rate, as seen experimentally in Ref.~\cite{Amasha2008} and anticipated theoretically in Ref.~\cite{Khaetskii2001} considering the Dresselhaus coupling only. In addition to the observed shift, the dot in-plane elongation is indicated also by our spectroscopy data (see Supplemental Material \cite{suppl}, Sec.~I): given the measured addition energy of about 3 meV, we would expect an orbital excitation energy of about 1 meV \cite{Hanson2007}, but in this sample, for the specific electrostatic configuration used for this experiment, the first orbital excitation energy is only 120 $\mu$eV.
We will therefore assume that the dot is strongly anisotropic (elongated), with the confinement potential major axis rotated away from $[100]$ by an angle $\delta$. Neither this angle, nor the degree of anisotropy (nor any more details on the potential shape) are known.

To derive an analogue of Eq.~\eqref{eq:rate 1} for an anisotropic dot, one should express Eq.~\eqref{eq:nso} in coordinates $x^\prime$, $y^\prime$, rotated from the crystallographic axes by the angle $\delta$, 
\begin{equation}
{\bf n}={\bf n}_{x^\prime} x^\prime + {\bf n}_{y^\prime} y^\prime.
\end{equation}
For an elongated dot, the excitations along the major axis ($x^\prime$) dominate the transition matrix element (see Supplemental Material \cite{suppl}, Sec.~II), and the rate is \cite{Khaetskii2001,Stano2006,Stano2006a}
\begin{equation} 
\Gamma \propto | {\bf B} \times {\bf n}_{x^\prime}|^2.
\label{eq:rate 2}
\end{equation}
After some trigonometric manipulations, we are able to write the previous equation in the form
\begin{equation}
\Gamma =b \left[ \sin^2 \xi + \cos^2 \xi \sin^2(\phi-\phi_{min}) \right],
\label{eq:elongated}
\end{equation}
where $b\equiv\kappa l_{so}^{-2}\left(1+ \sin 2\delta \sin 2\vartheta\right)$, with $\kappa$ a proportionality constant that sets the overall scale.

\begin{figure}
\includegraphics[width=8.5cm] {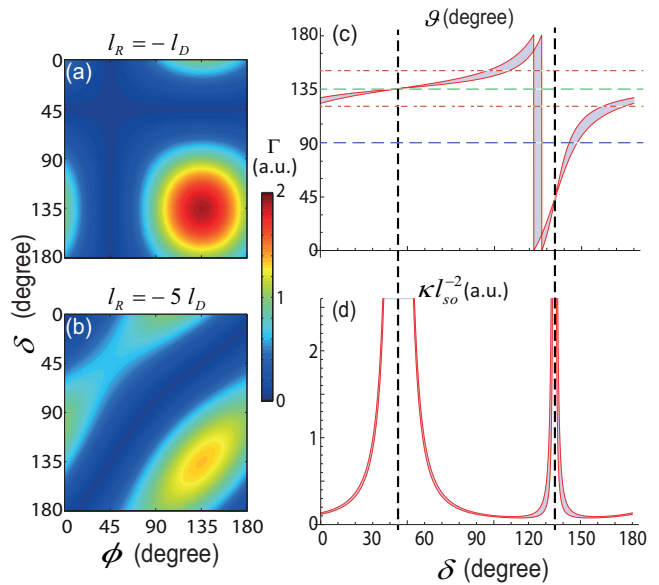}
\caption{\label{fig:fig4} Calculated values of $\Gamma$ (Eq.~\eqref{eq:elongated}) as a function of the angle of the external magnetic field $\phi$ and the dot major axis $\delta$, for $\xi=0$ and (a) $l_R/l_D=-1$ and (b) $l_R/l_D=-5$. (c-d) The result of the fit of the data of $1/T_1(\phi)$ from Fig.~\ref{fig:fig3}(b) to Eqs.~\eqref{eq:elongated} and \eqref{eq:angle}, with $\vartheta$ and ($\kappa l_{so}^{-2}$) (the latter in arbitrary units) as fit parameters, as a function of $\delta$. The shaded area between the two red curves indicates the $95\%$ confidence interval (not taking into account the systematic error in $\phi$). The two black vertical lines indicate $\delta=45^\circ,135^\circ$; the two red dotted lines and the green and blue horizontal lines are at $\vartheta=120^\circ,150^\circ$, $135^\circ$ and $90^\circ$ respectively.
}
\end{figure} 

This expression predicts a rate varying sinusoidally upon in-plane rotation of the magnetic field, a feature in common with Eq.~\eqref{eq:circular 1}. However, the details of the dependence are very different. Here, unlike in Eq.~\eqref{eq:circular 1}, the $\it{magic}$ magnetic field angle for which
the rate is minimal does depend on the ratio of Rashba and Dresselhaus coefficients (through $\vartheta$), and on the anistropy axis of the dot (through $\delta$):
\begin{equation}
\tan \phi_{min} = -\frac{\cos(\delta-\vartheta)}{\sin(\delta+\vartheta)}.
\label{eq:angle}
\end{equation}
To illustrate further the dependence of the relaxation rate on the orientation of the in-plane field and the dot major axis, we plot the prediction of Eq.~\eqref{eq:elongated} in Fig.~\ref{fig:fig4}(a-b) for different ratios of $l_R$ and $l_D$. When $l_R=-l_D$, the relaxation rate is minimal for $\phi=45^\circ$, regardless of the dot orientation, and also for $\delta=45^\circ$ regardless of the in-plane magnetic field orientation [Fig.~\ref{fig:fig4}(a)]. When $l_R \neq l_D$, the field orientation that minimizes the relaxation rate depends on the dot major axis orientation [Fig.~\ref{fig:fig4}(b)], with a $\pi$ periodicity.

We fit the data of Fig.~\ref{fig:fig3}(b) to Eq.~\eqref{eq:elongated} with $\phi_{min}$, $\xi$ and $b$ as the fitting parameters. The fit is plotted in Fig.~\ref{fig:fig3}(b) as the red line. It agrees excellently with the data (fit goodness $R^2\approx 0.99$) and gives $\phi_{min} =35.1^\circ \pm 1.1^\circ$, $\xi \approx 17.4^\circ \pm 1.1^\circ$, and $b=(139.2 \pm 3.5$) s$^{-1}$. The fitted $17^\circ$ misalignment of the magnetic field out of the plane is, however, unrealistically large. We estimated it in our experimental setup via Shubnikov-de Haas oscillations, and can put an upper limit $\left|\xi\right| < 5^\circ$ (see Supplemental Material \cite{suppl}, Sec.~I). The unexpectedly high value of $\xi$ comes from the large value of the relaxation rate at its minimum. We note, however, that this minimum value may also be dominated by other relaxation mechanisms that do not depend on $\phi$, such as the interaction with nuclear spins \cite{Khaetskii2001}, or the contribution from the random part of the R SO coupling \cite{Sherman2005} which arises due to fluctuations in the dopant concentration in the $\delta$-doping layer. Contributions of orbital excitations along the minor axis also lead to a finite offset, as is suggested by Eq.~\eqref{eq:rate 1}. Without knowing more about the dot confinement shape, we did not find it reasonable to try to separate these possible contributions by introducing more fitting parameters. Instead, we relabel $\xi\to \xi^*$, reinterpreting it as an effective angle accounting for all these possibilities together.

Using the value of $\xi^\ast$ obtained from the fit and Eqs.~\eqref{eq:elongated} and \eqref{eq:angle}, we can also perform a fit of the same data set with $\vartheta$ and $\kappa l_{so}^{-2}$ as free parameters, as a function of $\delta$. The fit results are plotted in Fig.~\ref{fig:fig4}(c-d). From there we conclude that without knowing the value of $\delta$, we can not establish the relative strength of the R and D couplings, as all values of $\delta$ are possible, in principle. However, we can infer that, most probably, in our sample $\alpha$ and $\beta$ were of comparable magnitude and opposite sign [$120^\circ\leq\left(\vartheta=\mbox{arctan}(l_r/l_d)\right)\leq150^\circ$], as these choices cover the larger portion of (a priori equally probable) values of $\delta$. There are two points, $\delta=45^\circ$ and $135^\circ$, where the rate $\kappa l_{so}^{-2}$ diverges (see Supplemental Material \cite{suppl}, Sec.~III). This indicates that such values of $\delta$ can not be reconciled with our data. Indeed, as follows from Eq.~\eqref{eq:angle}, for these values $\phi_{min}$ does not depend on the SO couplings, and should be $45^\circ$ or $135^\circ$, different from what we measured. We furthermore note that if $\delta$ were known, $\alpha/\beta$ could be extracted directly. In order to also determine the absolute values of $\alpha$ and $\beta$, more information is needed, such as the energy level spectrum of the dot. 

For future experiments, we give guidance for the optimal orientation of the quantum dot gate pattern and magnetic field relative to the crystal axes. First, since spin relaxation and EDSR based spin manipulation are governed by the same matrix elements for spin transitions, it is possible to simultaneously optimize for fast EDSR driven Rabi oscillations and for fast relaxation (useful for qubit reset \cite{Srinivasa2013}). In contrast, slow relaxation (useful for high-fidelity read-out \cite{Elzerman2004,Nowack2011}) cannot be optimized together with fast EDSR, as long as the phonon coupling is isotropic, as then both the spin relaxation rate and the EDSR rate scale with the same factor. In circular dots, this factor is given in Eq.~\eqref{eq:circular 2}. We see that the R and D terms maximally enhance or cancel each other when the external magnetic field is oriented along the  $[110]$ and $[1\overline{1}0]$ crystallographic axes, as can be expected also from Fig.~\ref{fig:fig1}(b). Complete cancellation of the two contributions is possible only when $|\alpha|=|\beta|$. When R and D have very different strengths, $\Gamma$ does not vary with the magnetic field orientation. For anisotropic dots, the factor is given in Eq.~\eqref{eq:elongated}. Here, $\Gamma$ oscillates with the field orientation and can reach zero (for $\xi=0$) regardless of the ratio of $\alpha$ and $\beta$. Finally, for maximizing the EDSR amplitude, in circular dots the external magnetic field has to point along $[110]$ ($[1\overline{1}0]$), if $\alpha\beta>0$ ($\alpha\beta<0$), and the driving electric field should be parallel to ${\bf B}$. In elongated dots the magnetic field should be oriented along the in-plane angle $\phi=\phi_{min}+\pi/2$, and the driving electric field should be along the dot soft axis. If the direction of the main dot axis can be chosen, it should point along $[110]$ ($[1\overline{1}0]$), if $\alpha\beta>0$ ($\alpha\beta<0$).
 
In conclusion, we show that the in-plane orientation of the magnetic field can strongly impact the spin relaxation time in quantum dots. We observe a variation in $T_1$ by more than an order of magnitude when rotating the field in the 2DEG plane. We can take advantage of this dependence in future experiments to either maximize or minimize $T_1$. Furthermore, the dependence of $T_1$ on magnetic field angle provides a sensitive probe of the ratio of the R and D SO contributions, which can be used even when SO induced avoided level crossings are too small to be measured \cite{Golovach2008}, which is the case of GaAs. What is needed is either a symmetric QD confining potential or, for an elliptical dot, a good estimate of the magnitude and direction of the QD anisotropy. Similar considerations are also valid for singlet-triplet qubits \cite{Kornich2014}\footnote{According to our analysis (not reported here), the dipole matrix elements between a singlet [either (1,1) and (0,2)] and a polarized triplet (1,1) in a biased two electron double dot tuned close to a singlet(1,1)-singlet(0,2) anti-crossing have the same angular dependence as the single electron matrix elements discussed here.}, where the easy axis is given by the double dot dipole axis \cite{Golovach2008}.\\

\begin{acknowledgments}
We acknowledge K. C. Nowack for fabricating the sample, V. Golovach for useful discussions, and R. Schouten for technical support. Research was supported by the Intelligence Advanced Research Projects Activity through the Army Research Office grant W911NF-12-1-0354, the European Research Council, the Dutch Foundation for Fundamental Research on Matter and the Swiss National Science Foundation.
\end{acknowledgments}

\appendix

\section{I. Experimental details}

\begin{figure}
\includegraphics[width=8.5cm] {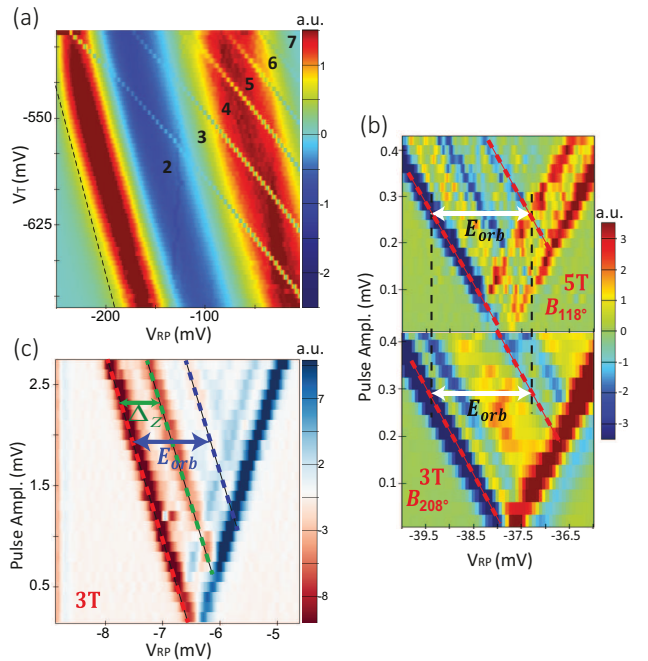}
\caption{\label{fig:sup1} (a) Derivative of the QPC current with respect to the gate voltage on RP, as function of the gate voltages on RP and T. The wide modulations (parallel to the black dotted line) are due to resonances in the QPC transport, which affect the QPC sensitivity. The sharp lines correspond to charge transitions in the dot. Usually, the region below the last transition is inferred to have zero electrons. From pulse spectroscopy data discussed below, we find that there are still two electrons left. (b) Pulse spectroscopy data \cite{Elzerman2004a} for two orientations of a 3 T field, ${\bf B}_{\phi}$, $90^\circ$ apart (c) Pulse spectroscopy measurement at 3 T at the gate voltage configuration used for the experiment. The green and blue dashed lines indicate the spin excited state and the first orbital excited state, respectively. The length of the blue arrow $(E_{orb})$ is comparable to twice of the length of the green $(\Delta_z)$ arrow.
}
\end{figure}

\begin{figure}
\includegraphics[width=8.5cm] {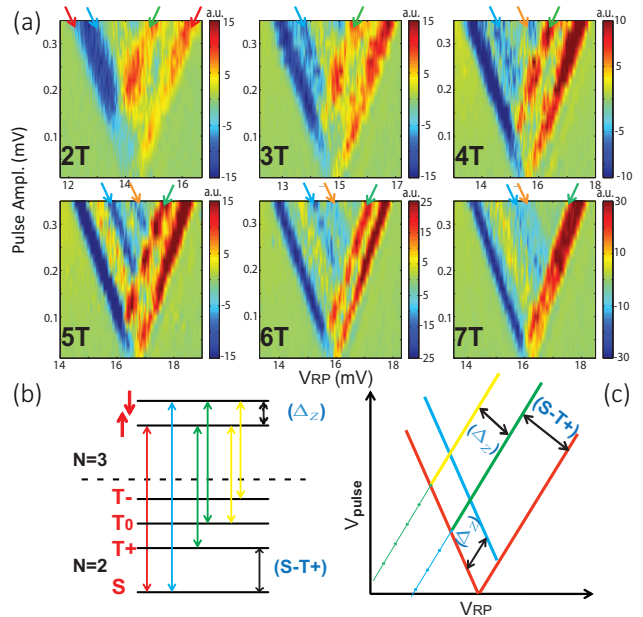}
\caption{\label{fig:sup2} (a) Pulse spectroscopy measurements \cite{Elzerman2004a} for different values of $\left|{\bf B}_{118^\circ}\right|$ (from 2 to 7 T in 1 T steps). The red and blue arrows indicate, respectively, the position of the $\uparrow$ (ground) and $\downarrow$ (excited) states of the ground orbital state, separated by the Zeeman energy $\Delta_z$. The green arrow denotes a two electron excited state (N=2). In particular, the distance (in mV on the x-axis) between this line and the right edge of the pulse-triangle represents the energy difference between the singlet (S) and $m=0$ triplet (T+), which are, respectively, the ground and first excited two electron states (see panel (b)). We find that the S-T+ energy splitting gets smaller with increasing magnetic field, with the same g-factor as the (N=3) orbital ground state Zeeman splitting. For 4, 5, 6, and 7 T we can distinguish an extra light blue line in the data, indicated by the orange arrow. We attribute it to the first orbital excited state of the system, $E_{orb}$ and we find that its distance from the left edge of the triangle is independent on $\left|{\bf B}_{118^\circ}\right|$. The gate voltage configuration used in this measurement is slightly different from the one used for the relaxation time measurement, giving a less elongated confining potential and a higher $E_{orb}$ than in in Fig.~\ref{fig:sup1}(c). (b) Schematic of the energy levels involved in the energy spectrum of the 2-3 electron charge transition \cite{Hanson2007}. Arrows with the same color indicate transitions between configurations with the same energy difference, which translates in a single line in spectroscopy measurements. (c) Schematic of the typical pulse-spectroscopy picture (red lines) resulting from an application of a square gate voltage pulse with increasing amplitude (y-axis), while stepping the DC voltage on the same gate (x-axis). The case is shown for the  2-3 electron charge transition. The color of each line corresponds to the transitions shown in (b). For the pulse amplitude window we used here only the $(T+ \rightarrow \uparrow)\equiv(T0\rightarrow \downarrow)$ transition is visible (the green line and arrows).
}
\end{figure}

\begin{figure}
\includegraphics[width=8.5cm] {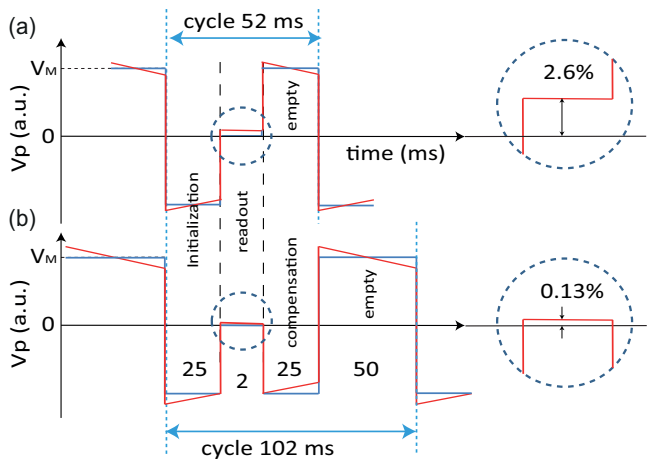}
\caption{\label{fig:pulse} The blue curves represent schematically the gate voltage pulse produced by the waveform generator for two pulse schemes. The red curves show the signal that arrives at the sample, after being distorted in the bias tee. (a) Three-stage pulse, keeping the length of the initialization-waiting stage and empty stage the same (see text). For waiting times comparable to the RC time constant of the bias-tee, the pulse gets significantly distorted by the charging of the bias-tee capacitance (inset on the right); this affects the stability of the read-out position. The deviation from the the ideal position ($V=0$) reaches  $2.6\%$ of the maximum pulse amplitude $V_M$ for a 25 ms initialization-waiting time. (b) Four-stage pulse scheme used to measure $T_1 > 10$ ms. The extra compensation stage (of the same length as the waiting time) reduces appreciably the deviation of the read-out position, to about $0.13\%$ of $V_M$ for a 25 ms waiting time, but almost doubles the total pulse time. The voltage deviations reported here have been estimated by a simulation of the bias-tee circuit made with Micro-cap.
}
\end{figure}

Here we give several experimental details, which were omitted in the main text due to space limitations.

The sample is mounted on a printed circuit board (PCB) which is attached via a coldfinger to the mixing chamber of a dilution refrigerator. The orientation of the sample with respect to the PCB is determined optically, with an estimated error of no more than $\pm3^\circ$. We can apply a magnetic field ${\bf B}_{\phi}$ in the 2DEG plane, at an angle $\phi$ with respect to the $[100]$ crystal axis, which can be controlled via a 2D vector magnet.
We tune the device to the few-electron regime [Fig.~\ref{fig:sup1}(a)] and adjust the tunnel couplings via the gate voltages. From analysing the pulse spectropy data of Fig.~\ref{fig:sup2} in detail, we conclude that the last transition seen in Fig.~\ref{fig:sup1}(a) is the transition between the two and three electron charge states. \\
The tunnel barrier between gate T and B is closed (tunnel rate $<$ 1 Hz). The barrier between gate T and RS is tuned to around 5 kHz. A coax line is connected to gate RP via a homemade resistive bias tee (R = 10 M$\Omega$, C = 47 nF, 1/RC $\approx$ 2 Hz) to allow fast pulsing of the dot levels while also maintaining a DC bias on RP [as indicated in Fig.~\ref{fig:fig1}(a)].\\
To measure the spin relaxation rate we apply a multi-stage voltage pulse to gate RP \cite{Elzerman2004}, using a Tektronix AWG5014. The simplest version of this pulse has three stages [blue line in Fig.~\ref{fig:pulse}(a)]. First we empty the QD by pulsing the ground state electrochemical potential above the lead Fermi level. A second pulse brings both the spin-up and spin-down levels below the lead Fermi energy, pulling one electron of unknown spin state into the QD. The last stage takes the dot to the read-out configuration, with the lead Fermi energy positioned in between the spin-up and spin-down levels. Here the electron tunnels out if and only if it is spin down. A tunnel event is reflected in the signal of the charge detector. Varying the initialization-waiting time between injection and read-out and monitoring the fraction of the time a tunnel event is seen (we typically average over 1000-5000 cycles), one can estimate the spin relaxation time, $T_1$, from the exponential decay of the measured spin down probability.\\
This three-stage pulse causes two potential artifacts when applied to the gate via the capacitor in the bias tee. First, if the pulse contains a DC component, it is blocked by the capacitor, thereby shifting the dot levels away from the desired configuration during read-out. Therefore we keep the average (DC) pulse amplitude fixed at zero, by compensating changes in the length or amplitude of the initialization-waiting stage by similar changes in the amplitude and length of the empty stage [see Fig.~\ref{fig:pulse}(a)].  Second, the high-pass filtering effect of the bias tee makes the pulse amplitude decay exponentially during every stages of the pulse, making the compensation less effective (red lines in Fig.~\ref{fig:pulse}). In order to further improve the stability of the gate voltage during read-out, we use the four-stage pulse schematically shown in Fig.~\ref{fig:pulse}(b), which introduces an extra compensation stage just after the read-out stage. This makes the alignment of the dot levels more independent from the waiting time, thereby reducing errors on the measured $T_1$.\\ 
A further experimental difficulty, which is most severe when applying the magnetic field along specific angles $\phi$, is the coupling in of mechanical vibrations into the measurement wires, possibly by magnetic flux induced currents in ground loops. It makes the dot potential oscillate relative to the Fermi level of the reservoir at the frequency of the mechanical vibration. This hinders the spin relaxation measurement and is the reason why we lack data points in some intervals of $\phi$ (e.g. $50^\circ<\phi<100^\circ$ in Fig.~\ref{fig:fig3}).\\
Finally, we evaluated the unintentional out-of-plane component of the applied magnetic field based on Shubnikov-de Haas oscillations, and estimate a misalignment of at most 5 degrees. We note that the out-of-plane component oscillates with the field orientation. In Fig.~\ref{fig:sup1}(b), we see no significant variation in $E_{orb}$ extracted from pulse spectroscopy when rotating the magnetic field over 90 degrees. This means that the small perpendicular magnetic field component will not significantly affect the measured spin relaxation times through its effect on the orbital level spacing. 

\section{II. Derivation of Eqs.~(2) and (6) of the main text}

The transition rate induced by phonons between the two lowest Zeeman split states, $\Psi_\uparrow$, $\Psi_\downarrow$, is in the lowest order of the electron-phonon interactions given by Fermi's golden rule, as
\begin{equation}
\Gamma=\frac{2\pi}{\hbar}\sum_{\alpha} |D_{\alpha}|^2 |R_{\alpha}|^2 \delta(\hbar \omega_\alpha - g\mu_B B).
\label{eq:spin relaxation rate}
\end{equation}
The sum is over acoustic phonons labeled by index $\alpha$ comprising phonon polarization and momentum, constrained by energy conservation requiring the phonon energy $\hbar \omega_\alpha$ to be equal to the energy splitting of the initial and final states, here the Zeeman energy $g \mu_B B$. Interested in the angular anisotropies for which an overall scale is unimportant, we do not specify the complex coefficients $D_\alpha$. In the dipole approximation of the phonon displacement operator $\exp(i {\bf k} \cdot {\bf r}) \approx 1+i{\bf k} \cdot {\bf r}$ (here ${\bf r}$ is the electron coordinate and ${\bf k}$ is the inplane phonon momentum), $R$ is a dipole matrix element 
\begin{equation}
R = 
\langle \Psi_{\downarrow} | {\bf k}\cdot {\bf r} | \Psi_{\uparrow} \rangle,
\label{eq:relaxation matrix element}
\end{equation}
which is non-zero only due to the SO interactions. Treating them perturbatively, (see, e.g., Eq.~(A8) in \cite{Raith2012}) we obtain
\begin{equation}
R =
2\sum_{j\neq 0,s} E_{0j}^{-1} \langle \psi_{0\uparrow} |\mu {\bf B}_{\rm eff} \cdot \boldsymbol{\sigma}| \psi_{js}\rangle \langle \psi_{js} | {\bf k}\cdot {\bf r} |\psi_{0\downarrow} \rangle,
\label{eq:matrix element estimate a} 
\end{equation}
where the sum goes over the orbital excited states $\psi_{js}$ offset from the ground state $j=0$ by orbital excitation energies $E_{0j}$ and $s=\uparrow, \downarrow$ is the spin with the quantization axis along the external magnetic field. The states $\psi_{js}$ are those of a system without SO interactions, so that they are separable into an orbital part $|j\rangle$ and a spinor part $|\xi_s\rangle$. We denoted $\psi_{0s}$ as the state which develops into $\Psi_s$ appearing in Eq.~\eqref{eq:relaxation matrix element} upon adiabatically turning on the SO interactions.
Finally, the effective magnetic field is \cite{Baruffa2010}
\begin{equation}
\mu {\bf B}_{\rm eff} = \mu ({\bf n}\times {\bf B})\cdot \boldsymbol{\sigma},
\end{equation}
with the SO dependent vector
\begin{equation}
{\bf n} = x \left( l_{d}^{-1}, -l_{r}^{-1},0\right) + y \left( l_{r}^{-1}, -l_{d}^{-1},0 \right) = {\bf n}_x x + {\bf n}_y y, 
\label{eq:effective SO vector}
\end{equation}
where the last equality sign is a definition of ${\bf n}_{x,y}$, two in-plane vectors. 

Before evaluating for specific cases we simplify the squared dipole element to
\begin{equation}
\begin{split}
|R|^2 =& 4\sum_{j j^\prime\neq 0} \langle 0 |\mu {\bf B}_{\rm eff} | j \rangle \cdot \langle {j^\prime} |\mu {\bf B}_{\rm eff} | 0 \rangle \frac{\langle 0 | {\bf k}\cdot {\bf r} |j \rangle \langle {j^\prime} | {\bf k}\cdot {\bf r} |0 \rangle}{E_j E_{j^\prime}}.
\end{split}
\label{eq:matrix element estimate b} 
\end{equation}
In going from Eq.~\eqref{eq:matrix element estimate a} to Eq.~\eqref{eq:matrix element estimate b} we used that the effective magnetic field is perpendicular to the external magnetic field, and that the phonon dipole operator is diagonal in the spin space.

Consider now the case of a circularly symmetric dot. The orbital states can be labeled by the orbital momentum index $l$. Though we allow for a possible out-of-plane magnetic field, breaking the time reversal symmetry, we assume its orbital effects are not so strong as to cause state crossings compared to the zero magnetic field case case [see also the data of Fig.~\ref{fig:sup1}(b)]. This restriction is not essential for the results and we adopt it only to simplify the notation. Under this assumption, the orbital ground state is fully symmetric, $l=0$. The lowest two excited states are $l=\pm 1$, and are degenerate if the out-of-plane magnetic field is zero and energy split otherwise. We now approximate the sum over the whole spectrum by these two lowest excited states in Eq.~\eqref{eq:matrix element estimate b}. The circular symmetry of the Hamiltonian, and consecutively its eigenstates, allows us to derive
\begin{subequations}
\begin{eqnarray}
\langle 0 | x |+1\rangle \langle +1 | y | 0\rangle &=& - \langle 0 | y |+1\rangle \langle +1 | x | 0\rangle,
\label{eq:aux 1}\\
\langle 0 | x |+1\rangle \langle -1 | y | 0\rangle &=& - \langle 0 | x |-1\rangle \langle +1 | y | 0\rangle,
\label{eq:aux 2}\\
\langle 0 | x |+1\rangle \langle -1 | x | 0\rangle &=& - \langle 0 | x |-1\rangle \langle +1 | x | 0\rangle,
\label{eq:aux 3}
\end{eqnarray}
\label{eq:aux}
\end{subequations} 
which follow upon inserting the identity in the form of $R_{\pi/4} R_{-\pi/4}$ into Eq~\eqref{eq:aux 1} and $I_y I_y$ into Eqs.~\eqref{eq:aux 2} and \eqref{eq:aux 3}, with the operator of an in-plane rotation $R_{\alpha}|l\rangle = \exp(i \alpha l)$, and the inversion along the $y$ axis $I_y |l\rangle = |-l\rangle$, where we adopted a phase convention $\langle {\bf r} | +1 \rangle=\langle {\bf r} | -1 \rangle^\dagger$. With the auxiliary results in Eq.~\eqref{eq:aux}, we see that cross terms, such as $j\neq j^\prime$ and ${\bf B} \times {\bf n}_{x^\prime} \cdot {\bf B} \times {\bf n}_{y^\prime}$, cancel and Eq.~\eqref{eq:matrix element estimate b}, restricted to the lowest excited subspace contributions, takes the form
\begin{equation}
\begin{split}
|R|^2 =& \left( | \mu {\bf B}\times {\bf n}_x|^2 + | \mu {\bf B}\times {\bf n}_y|^2 \right) \times\\
&\qquad \sum_{j = \pm 1} 4 E_j^{-2} \langle 0 | x | j \rangle|^2 |\langle 0 | {\bf k}\cdot {\bf r} |j \rangle|^2,
\end{split}
\label{eq:matrix element estimate c} 
\end{equation}
which leads to Eq.~\eqref{eq:rate 1} of the main text.

The derivation for the case of an anisotropic dot is even simpler. Indeed, for such a dot there is a single lowest orbital excited state, $j=1$, dominating the sum in Eq.~\eqref{eq:matrix element estimate b}. The dipole matrix element of this exited state with the ground state is an in-plane vector ${\bf d} =\langle 0 | {\bf r} | 1\rangle$, which by its definition fulfills $\langle 0 | {\bf r}\cdot (\hat{z}\times {\bf d}) | 1\rangle=0$. Defining the rotated coordinated system with axes $x^\prime$ along ${\bf d}$, and $y^\prime$ perpendicular to it (say along $\hat{z}\times {\bf d}$), we immediately get
\begin{equation}
\begin{split}
|R|^2 =& | \mu {\bf B}\times {\bf n}_{x^\prime}|^2 \times 4 E_1^{-2} |{\bf d}|^2 |{\bf d}\cdot {\bf k}|^2,
\end{split}
\label{eq:matrix element estimate d} 
\end{equation}
which gives Eq.~\eqref{eq:rate 2} of the main text.

\section{III. Divergences in fitting parameter $\kappa l_{so}^{-2}$}

\begin{figure}
\includegraphics[width=8.5cm] {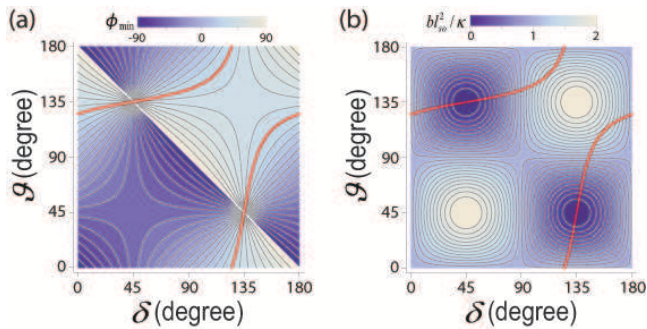}
\caption{\label{fig:sup3} Plots of (a) $\phi_{min}(\vartheta,\delta)$ [Eq.~\eqref{eq:angle}] and (b) $(1+\sin 2\vartheta \sin 2\delta)$ [in Eq.~\eqref{eq:elongated}] in the $(\vartheta,\delta)$ plane. The grey contour lines show $\phi_{min}=\chi$, with $-90^\circ\leq\chi\leq+90^\circ$ in 25 steps. The red curve represents the relation $\phi_{min}=35.1^\circ$ obtained from the fit to Eq.~\eqref{eq:elongated}. We notice that all the contour lines in panel (a) cross in the points $(45^\circ,135^\circ)$ and $(135^\circ,45^\circ)$ which are zeros of the angular part of the pre-factor $bl_{so}^{2}/\kappa$ (panel (b)). 
}
\end{figure}

The goal of this paragraph is to provide intuition for the presence and the shape of the two divergences, for $\delta=45^\circ,135^\circ$, in the fitting parameter $\kappa l_{so}^{-2}$ reported in Fig.~\ref{fig:fig4}(d) of the main text. Fitting the data of Fig.~\ref{fig:fig3}(b) to Eq.~\eqref{eq:elongated} we get $(b,\phi_{min},\xi^\ast)=(139.2\pm 3.5\;s^{-1},\;35.1^\circ\pm1.1^\circ,\;17.4^\circ\pm1.5^\circ)$. A plot of $\phi_{min}(\delta,\vartheta)$ (from Eq.~\eqref{eq:angle}) for $\delta,\vartheta\in[0^\circ,180^\circ]$ is presented in Fig.~\ref{fig:sup3}(a), together with a plot of the angular part of the pre-factor $b$, $(1+\sin2\delta \sin2\vartheta)=b l_{so}^2/\kappa$, in Fig.~\ref{fig:sup3}(b). The red curve on top of those two plots represents the condition $\phi_{min}(\delta,\vartheta)=35.1^\circ$, the value for the $magic$ angle obtained from Fig.~\ref{fig:fig3}. We notice that $(1+\sin2\delta \sin2\vartheta)=0$ for the coordinates $(\delta,\vartheta)=(45^\circ,\;135^\circ)$ and $(135^\circ,\;45^\circ)$, and that all the contour lines in Fig.~\ref{fig:sup3}(a) cross these points. In order to keep $b$ fixed to $139.2\pm 3.5\;s^{-1}$ (the value from the fit), the quantity $\kappa l_{so}^{-2}$ [plotted as a function of $\delta$ in Fig.~\ref{fig:fig4}(d)] should diverge around those two points of the $(\delta,\vartheta)$ plane. Furthermore, how fast $\kappa l_{so}^{-2}$ diverges is determined by the derivative with respect to $\delta$ along the red curve in Fig.~\ref{fig:sup3}(b) around the singularity points. This explains why in Fig.~\ref{fig:fig4}(d) the singularity around $\delta=135^\circ$ looks much sharper than the one around $\delta=45^\circ$.\\

%

\end{document}